\documentclass[slac_one]{revtex4}
\usepackage{graphicx}
\usepackage{fancyhdr}
\usepackage{xspace}
\usepackage{subfigure}

\begin{document}

\def\sla#1{\rlap{\kern .15em /}#1}
\newcommand{\pt}{\mbox{$p_T$}\xspace}
\newcommand{\Ht}{\mbox{$H_T$}\xspace}
\newcommand{\et}{\mbox{$E_T$}\xspace}
\newcommand{\met}{\mbox{$\sla{E}_T$}\xspace}
\newcommand{\eoverp}{\mbox{$E/p$}\xspace}
\newcommand{\isoetcorr}{\mbox {$E_ {T}^{Iso(corr)}$\xspace }}
\newcommand{\phojets}{\mbox{$\gamma$ + jets}\xspace}
\newcommand{\phoonejet}{\mbox{$\gamma$ + $\geq$1 jet}\xspace}
\newcommand{\photwojet}{\mbox{$\gamma$ + $\geq$2 jets}\xspace}
\newcommand{\phojetsmet}{\mbox{\phojets + \met}\xspace}

\title{Search for Anomalous Production of Photon + Jets} 

%

\author{J.~R.~Dittmann, S.~Hewamanage, N.~Krumnack}
\affiliation{Baylor University, Waco, TX 76798-7316, USA}
\author{R.~Culbertson, A.~Pronko}
\affiliation{FNAL, Batavia, IL 60510, USA}

\author{(On  behalf of the CDF Collaboration)}
\noaffiliation

\begin{abstract}
Many new physics models predict mechanisms that could produce a \phojets signature.  We search in the
\phojets channel, independent of any model, for new physics using 2~fb$^{-1}$  of CDF Run II data collected at the Fermilab Tevatron from $p\bar{p}$ collisions
at $\sqrt{s} = 1.96$ TeV. A variety of techniques are applied to estimate the Standard Model expectation and non-collision backgrounds. We examine several
kinematic distributions including photon $E_T$, invariant masses, and total transverse energy in the event for discrepancies with predictions of the Standard Model. 
\end{abstract}

\maketitle

\thispagestyle{fancy}
\section{INTRODUCTION}
\begin{figure*}[h!]
\centering
\subfigure[]{\label{fig-sm-feyn}
\begin{minipage}[b]{0.3\textwidth}
\includegraphics[scale=0.3]{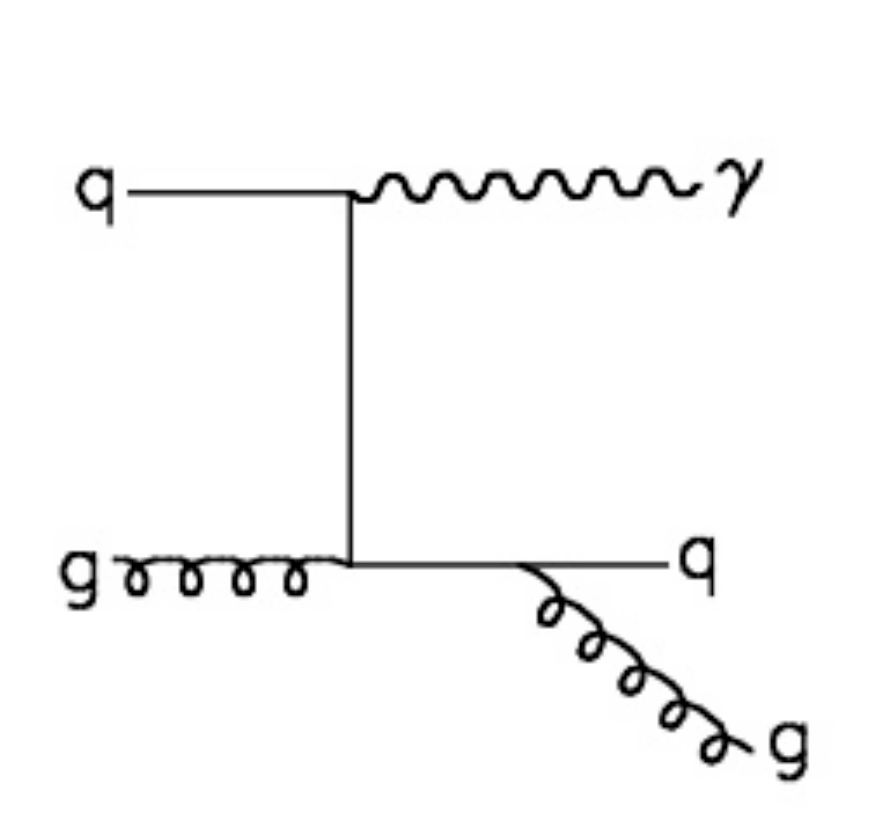}
\includegraphics[scale=0.3]{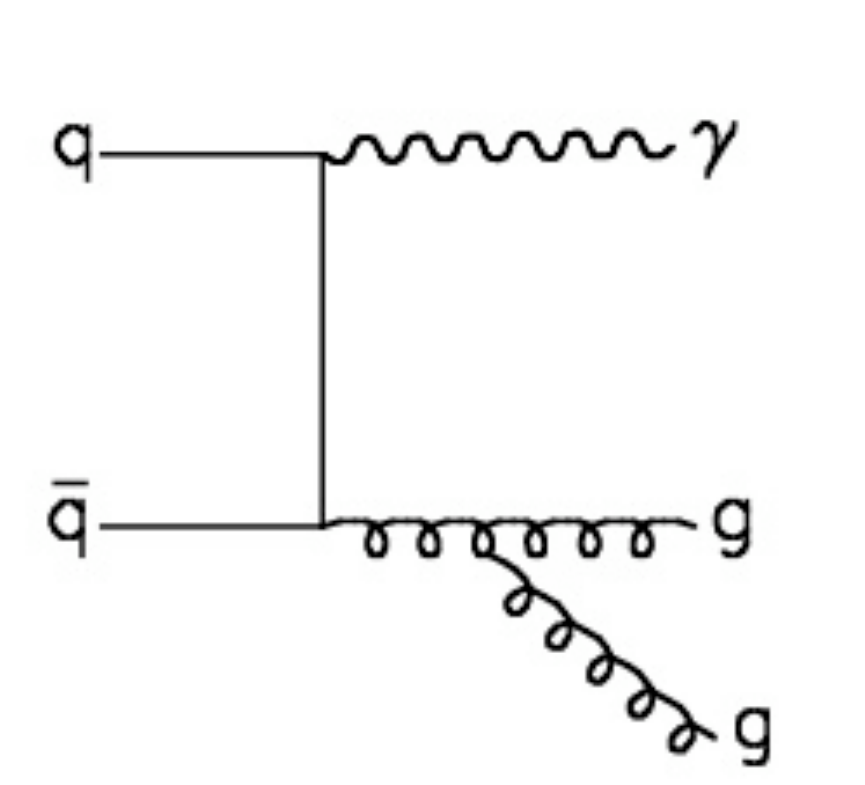}
\includegraphics[scale=0.3]{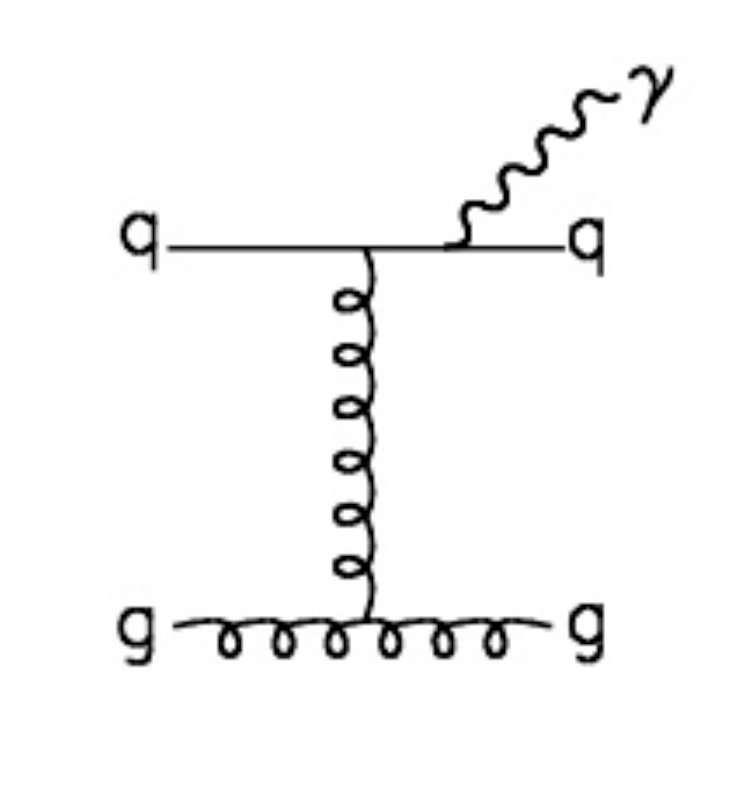}
\end{minipage}}
\qquad\qquad
\subfigure[]{\label{fig-gmsb-feyn}
\includegraphics[scale=0.28]{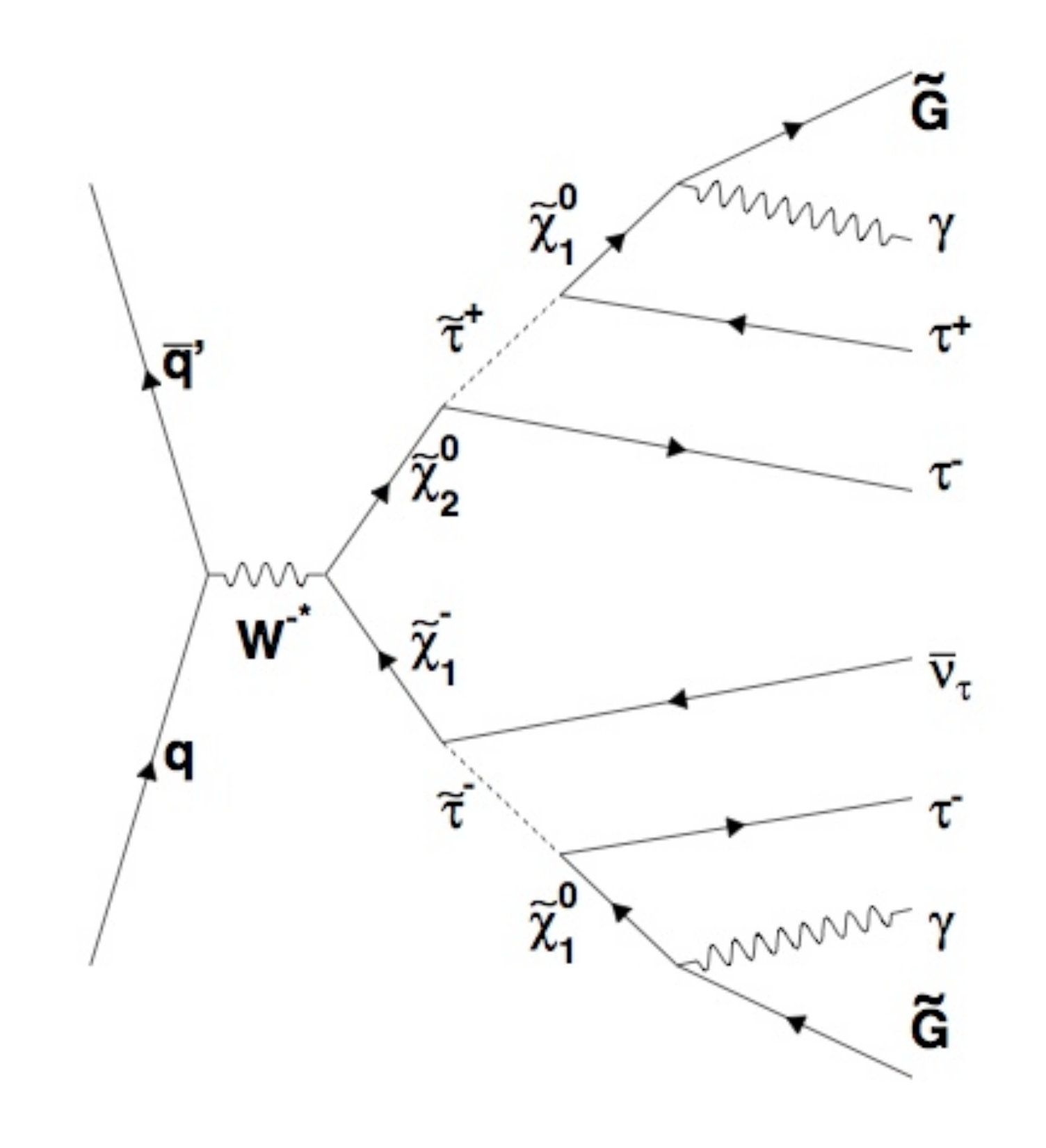}\quad
\includegraphics[scale=0.28]{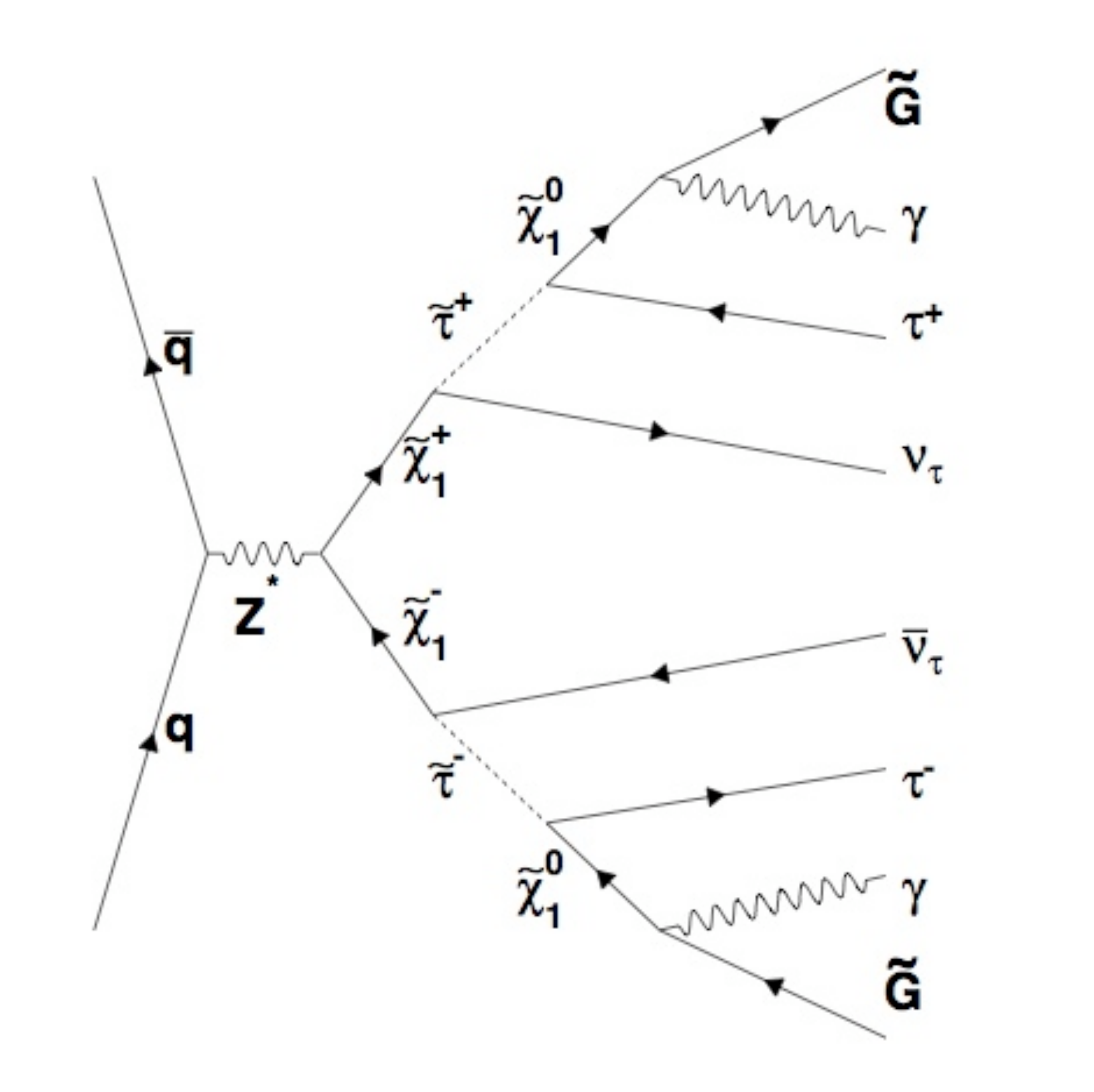}}
\caption{Feynman diagrams for tree-level (a) Standard Model and (b) GMSB processes that yield the $\gamma$~+~2 jets signature.}
\end{figure*}

We present the preliminary findings of a model-independent, signature-based search for the anomalous production of \phojets in $p\bar{p}$ collisions at
$\sqrt{s} = 1.96$ TeV using 2~fb$^{-1}$  of CDF Run II data. We scan kinematic distributions including photon ($\gamma$) energy, invariant mass of the
$\gamma$ and leading jets, and total transverse energy in the event ($H_{T}$) for an excess of events over Standard Model (SM) predictions.  An excess
could indicate the existence of a new heavy particle decaying into \phojets or a new physics mechanism such as gauge-mediated SUSY
breaking~\cite{ref-models}. The Feynman diagrams in Fig.~\ref{fig-sm-feyn} and \ref{fig-gmsb-feyn} illustrate examples of processes that yield
the $\gamma$~+~2 jets signature.

\section{SELECTING $\gamma$ + JETS EVENTS}
\begin{figure*}[t]\label{fig-evd}
\centering
\subfigure[\mbox{ A $\gamma$ + 1 jet event}]{\label{fig-p1j-edited}
\includegraphics[scale=0.2]{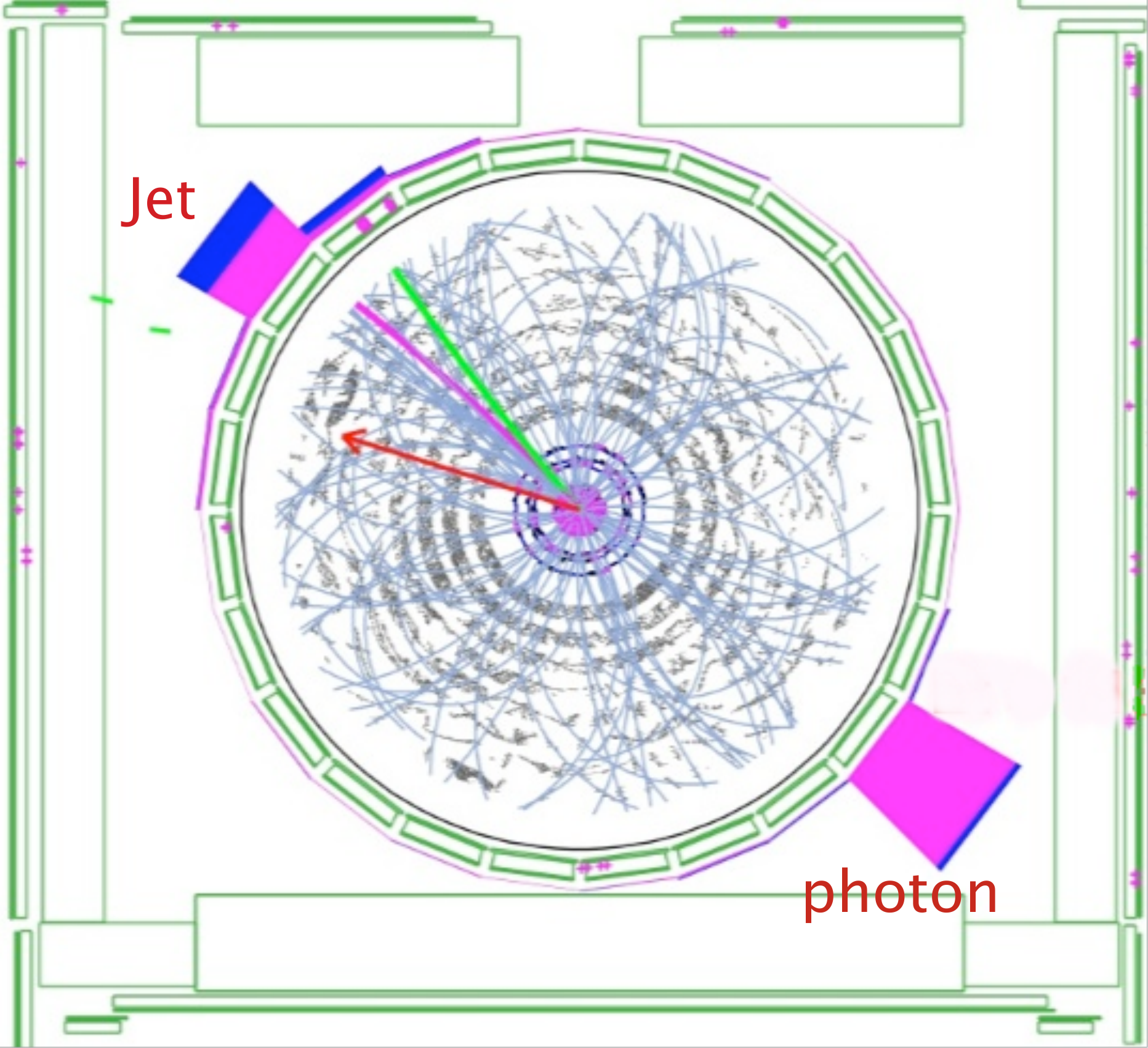}}
\qquad\qquad
\subfigure[\mbox{ A cosmic ray event}]{\label{fig-p1j-cosmic}
\includegraphics[scale=0.2]{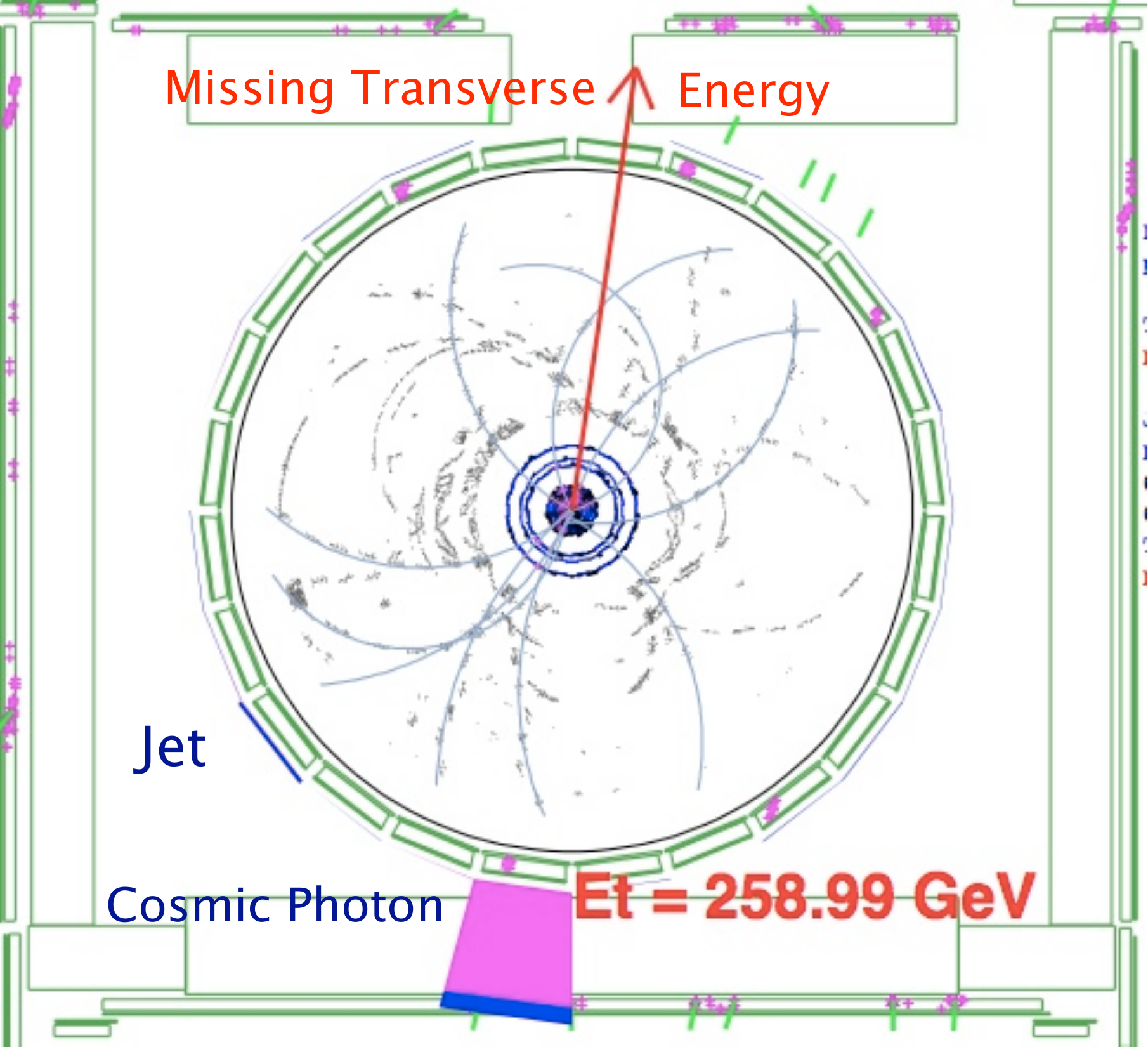}}
\caption{(a) Cross-sectional view of a $\gamma$ + 1 jet event in the CDF detector.  The $\gamma$ is opposite of the jet in the plane perpendicular to the beamline, hence there is no missing transverse energy.  (b) A typical cosmic ray event. There is little activity in the detector except for a large EM energy cluster, which mimics a $\gamma$, and just enough energy in the calorimeter to form a jet, making a perfect $\gamma$ + 1 jet signature. This topology will give rise to a large \met.}
\end{figure*}

We select a sample of $\gamma$ candidates by identifying electromagnetic (EM) clusters with transverse energy
\et~$>$~30~GeV in the central region of the calorimeter ($|\eta|<1.1$).  To reduce the background from charged leptons, we require an absence of tracks pointing in the direction of the EM cluster.   Background from cosmic rays is reduced with a requirement on calorimeter timing, and we remove events that originate from the beam halo using a set of topological selection requirements.  Events with photomultiplier tube spikes --- an instrumentation effect that can resemble a $\gamma$ ---  are also removed. In the remaining event sample, we identify one or more jets with \mbox{\et$>15$ GeV} and $|\eta| < 3.1$. A typical $\gamma$ + jet event in the CDF detector is depicted in Fig.~\ref{fig-p1j-edited}.

\section{MODELING BACKGROUNDS}
\begin{figure*}[tbh]\label{fig-backgrounds}
\centering
\subfigure[\mbox{ }An electron faking a $\gamma$]{\label{fig-electron-brem}
\includegraphics[scale=0.14]{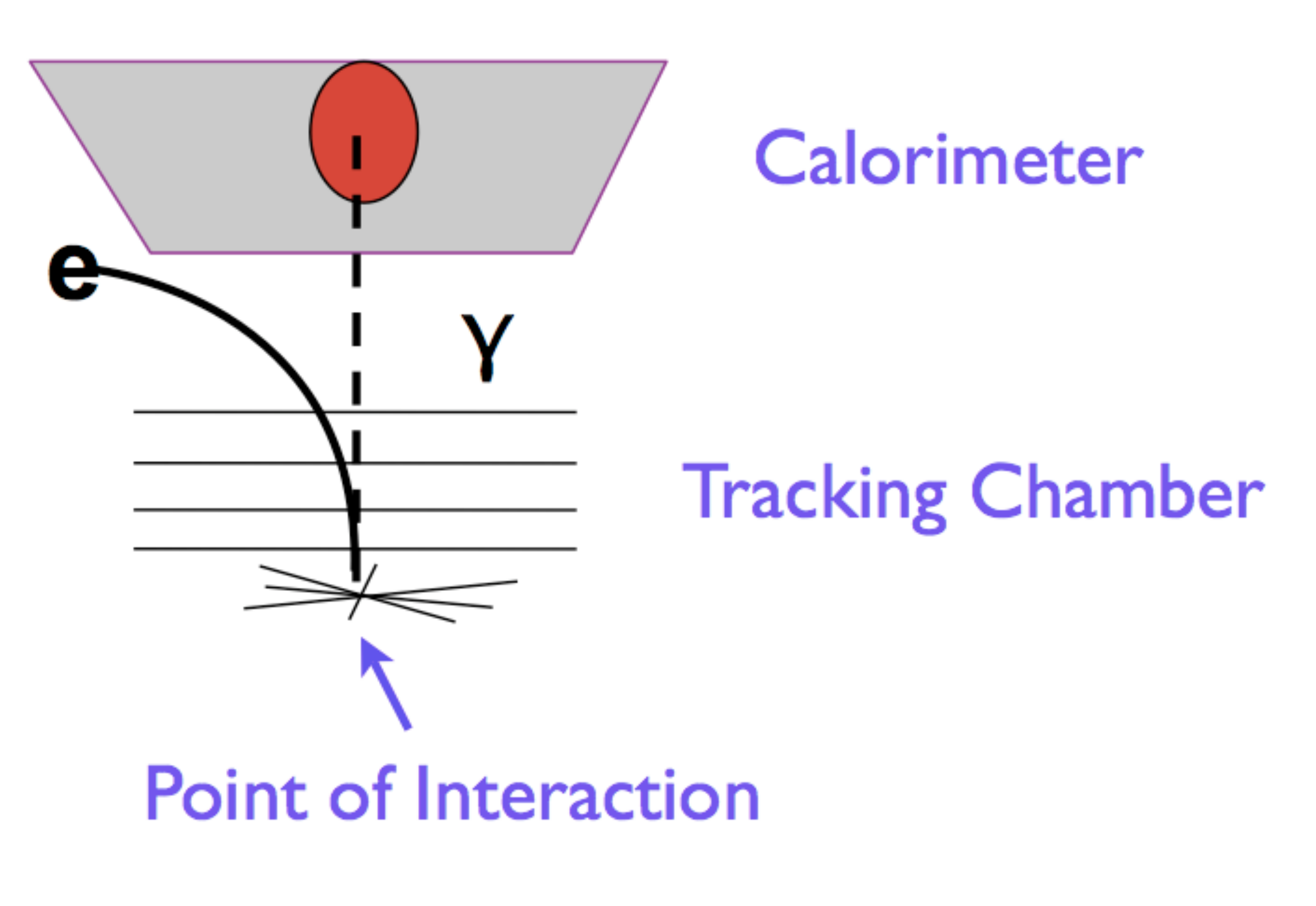}}
\qquad\qquad
\subfigure[\mbox{ }Beam halo photons]{\label{fig-beamhalo}
\includegraphics[scale=0.19]{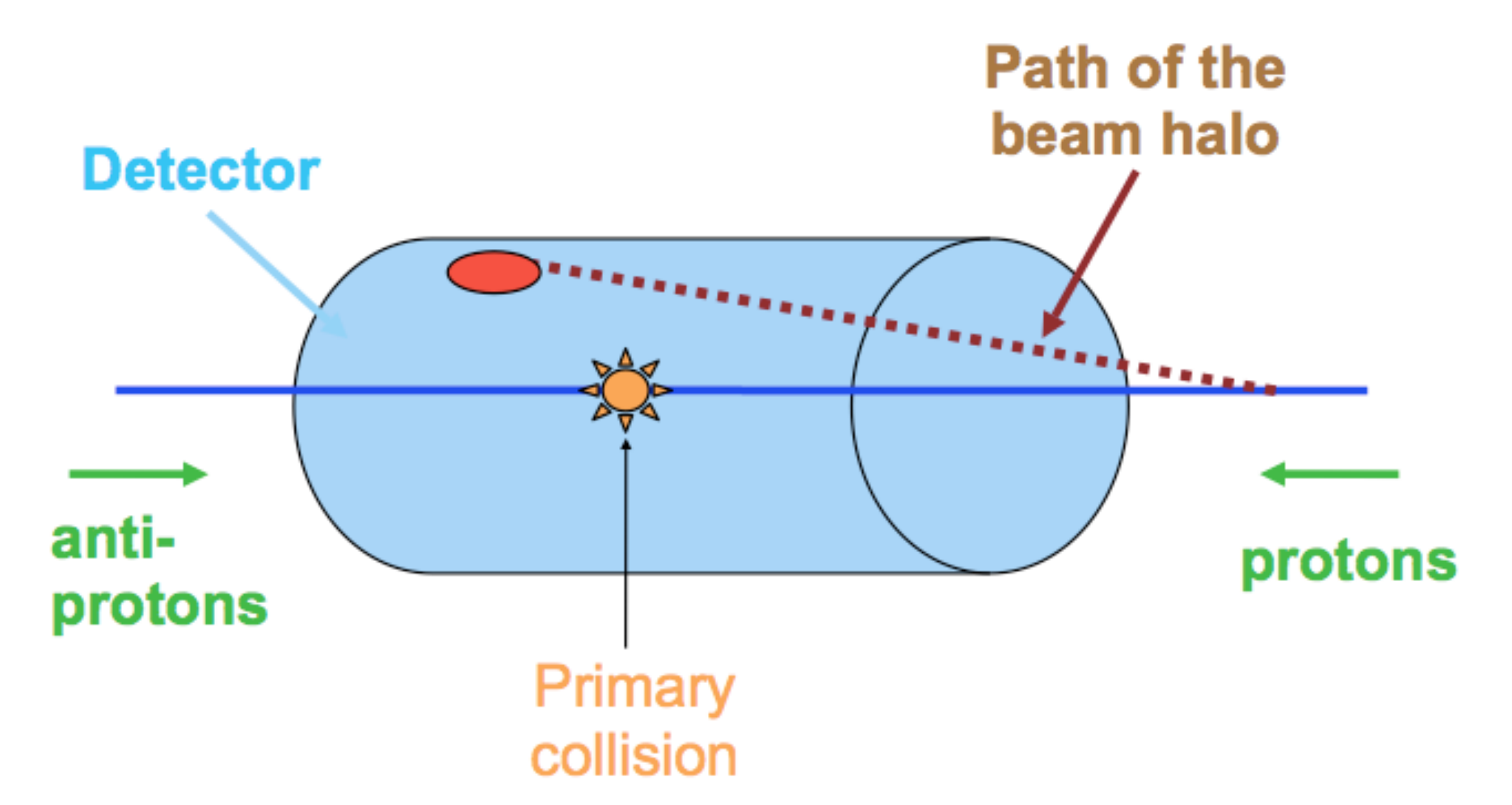}}
\qquad
\subfigure[\mbox{ }Calorimeter EM timing of photons]{\label{fig-calorimeter-emtiming}
\includegraphics[scale=0.14]{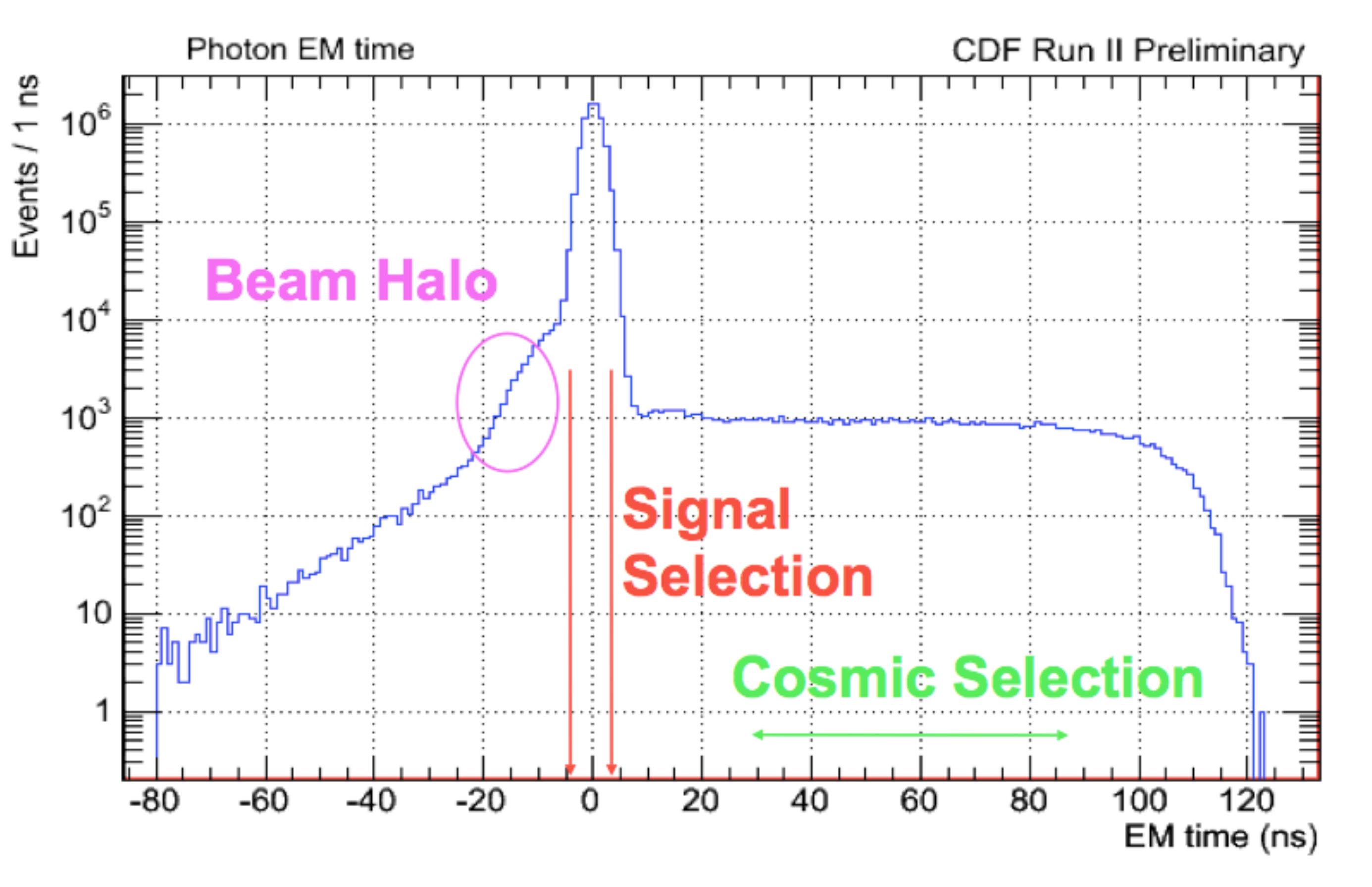}}
\caption{(a) An electron radiating a $\gamma$ loses momentum and drifts away from its original path in the magnetic field. Few hits are observed in the tracking chamber, and a $\gamma$ is reconstructed. (b) Fake photons from the beam halo. (c) Energy from the beam halo is detected earlier than photons from the primary collision, which generally arrive within \mbox{$|t|<4.8$ ns}. The time-independent cosmic photons are the flat component visible in the region \mbox{$t>10$ ns}.}
\end{figure*}

The SM $\gamma$ and SM charged lepton backgrounds are modeled using the \textsc{Pythia} Monte Carlo generator (Tune~A) \cite{ref-pythia}.
All other backgrounds are modeled using data. Background from QCD multijet production, in which a jet fakes a photon, are modeled using
a sample that consists of jets that pass the photon selection requirements.  Although a large portion of background from cosmic rays and the
beam halo is removed by the \phojets selection requirements, some events remain, and these backgrounds are significant in the large \met region.
A pure cosmic ray event sample is attained using EM timing information and is used to construct a background template.  A set of topological cuts
is used to select beam halo events. The beam halo and cosmic ray templates are normalized to the expected number of background events
in the \phojets sample. The SM lepton template is normalized according to the luminosity of the data. After subtracting all other backgrounds, the SM
$\gamma$ and multijet backgrounds are normalized according to the fake $\gamma$ fraction which is determined to be \mbox{0.319 $\pm$ 0.001(stat) $\pm$ 0.0068(syst)} for photons with \mbox{\et$>$ 30 GeV}.  Figure~\ref{fig-backgrounds} illustrates a few of these backgrounds, and
Table~\ref{tab-bg-estimates} summarizes the background estimates.

\begin{table}[ht]
\begin{center}
\caption{Summary of background estimates. Standard Model photons and multijet backgrounds are dominant by several orders
of magnitude. Setting a limit on \met in the event will reduce these dominant backgrounds and make others significant.}
\begin{tabular}{|l|c|c|}
\hline \textbf{Background} & \textbf{Events Expected for $\gamma~+ \geq$ 1 Jets} & \textbf{Events Expected for $\gamma~+ \geq$ 2 Jets} \\
\hline SM Photons & 2.6M & 650K \\
\hline QCD Multijet	 & 1M & 280K \\
\hline SM Leptons & 5K & 1K \\
\hline Cosmic Ray Photons & 110 & 7 \\
\hline Beam Halo Photons	 & 9 & $\le$1\\
\hline Photons from PMT Spikes & 0 & 0 \\
\hline
\end{tabular}
\label{tab-bg-estimates}
\end{center}
\end{table}

\section{PRELIMINARY RESULTS}
We present a few of the $\gamma$ + $\geq$1 jet results in Fig.~4. The data are represented by black dots and backgrounds are shown in different colors. The shaded region signifies the total systematic uncertainty. Uncertainty due to the jet energy measurement is by far the largest systematic uncertainty. Uncertainties in determining the fake $\gamma$ fraction, integrated luminosity, EM energy measurements, beam halo estimate, and cosmic ray background estimate are also taken into account. We have measured the photon \et spectrum from \mbox{30 GeV} to about \mbox{550 GeV}, and over this range the total systematic uncertainty increases from 15\% to 90\%. It is evident that at higher \et the photon purity increases.  We are limited by statistics at high \et. The invariant mass of the $\gamma$ and the leading jet extends up to \mbox{1000 GeV}. Many background predictions become limited by statistics in the high mass region, and the systematic uncertainty increases from 15\% to 90\%. As is evident from these plots, the SM $\gamma$ and QCD multijet  backgrounds are dominant.

An excess of data over the background prediction and uncertainties will hint at new physics. Thus far, we see good agreement with the present theory predictions extending over several orders of magnitude. In the final stage of this analysis we plan to perform a closer inspection of events with large \met.

\begin{figure*}[tbh]\label{fig-allplots}
\centering
\rotatebox{90} {\includegraphics[scale=0.3]{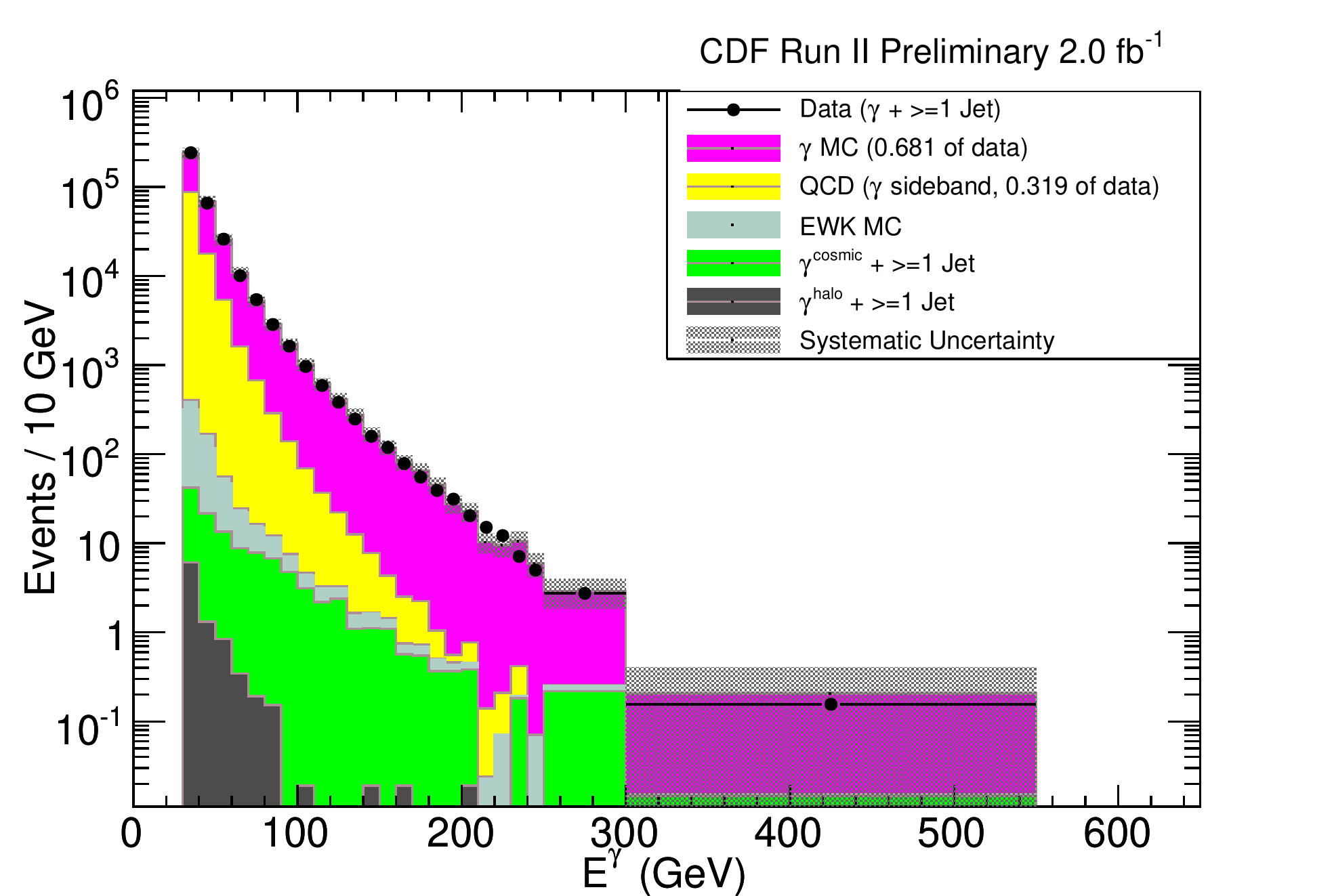}}
\rotatebox{90} {\includegraphics[scale=0.3]{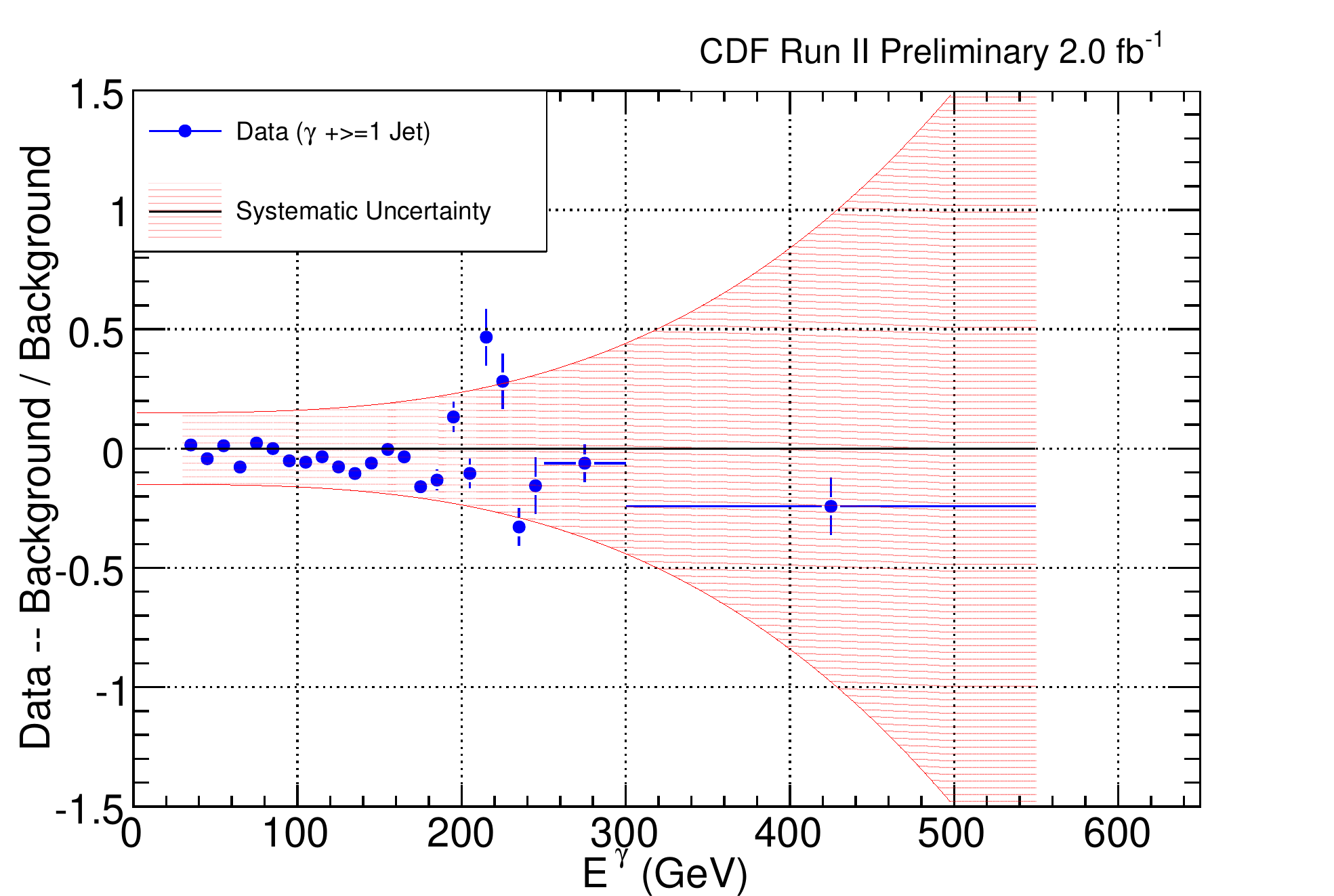}}
\rotatebox{90} {\includegraphics[scale=0.3]{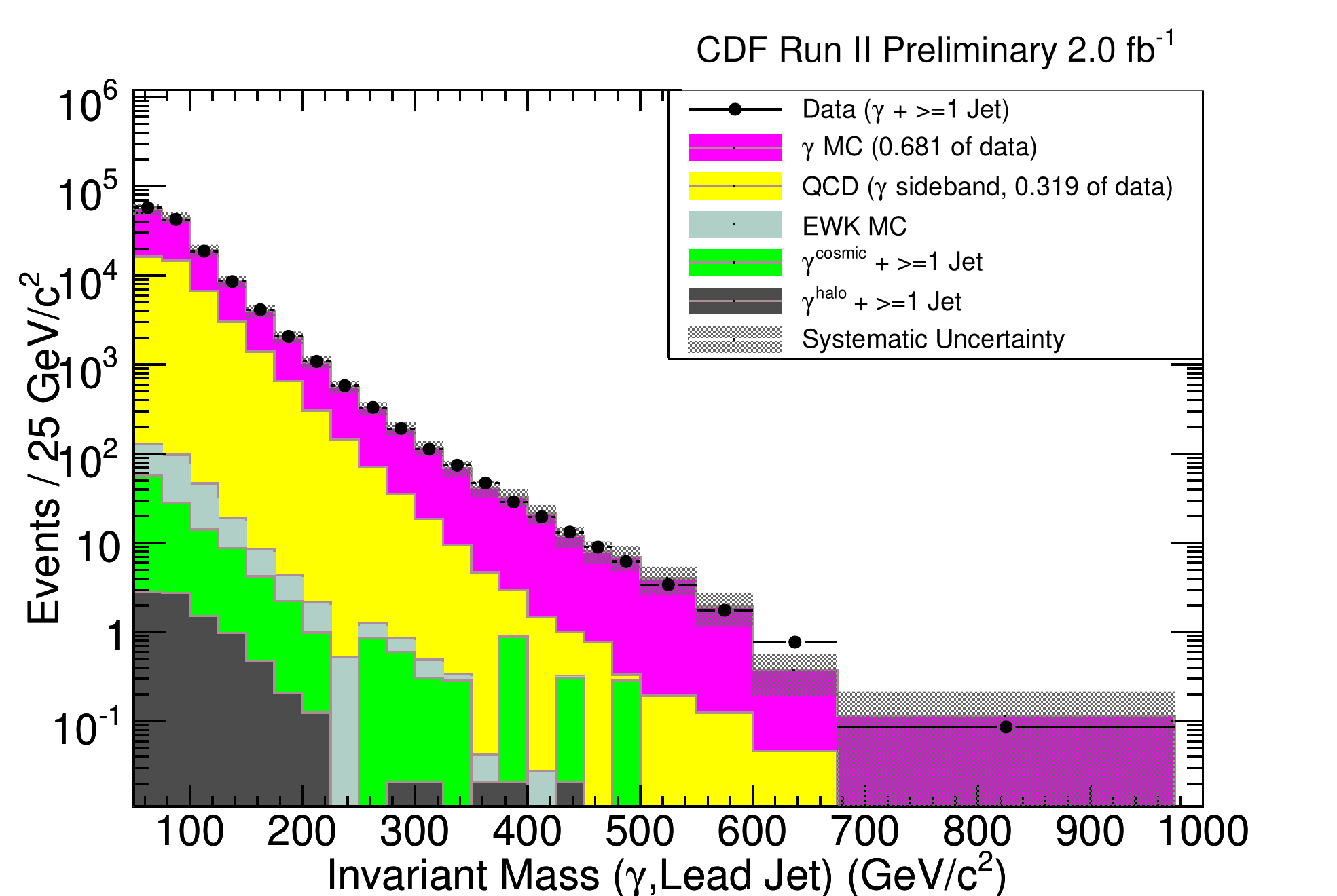}}
\rotatebox{90} {\includegraphics[scale=0.3]{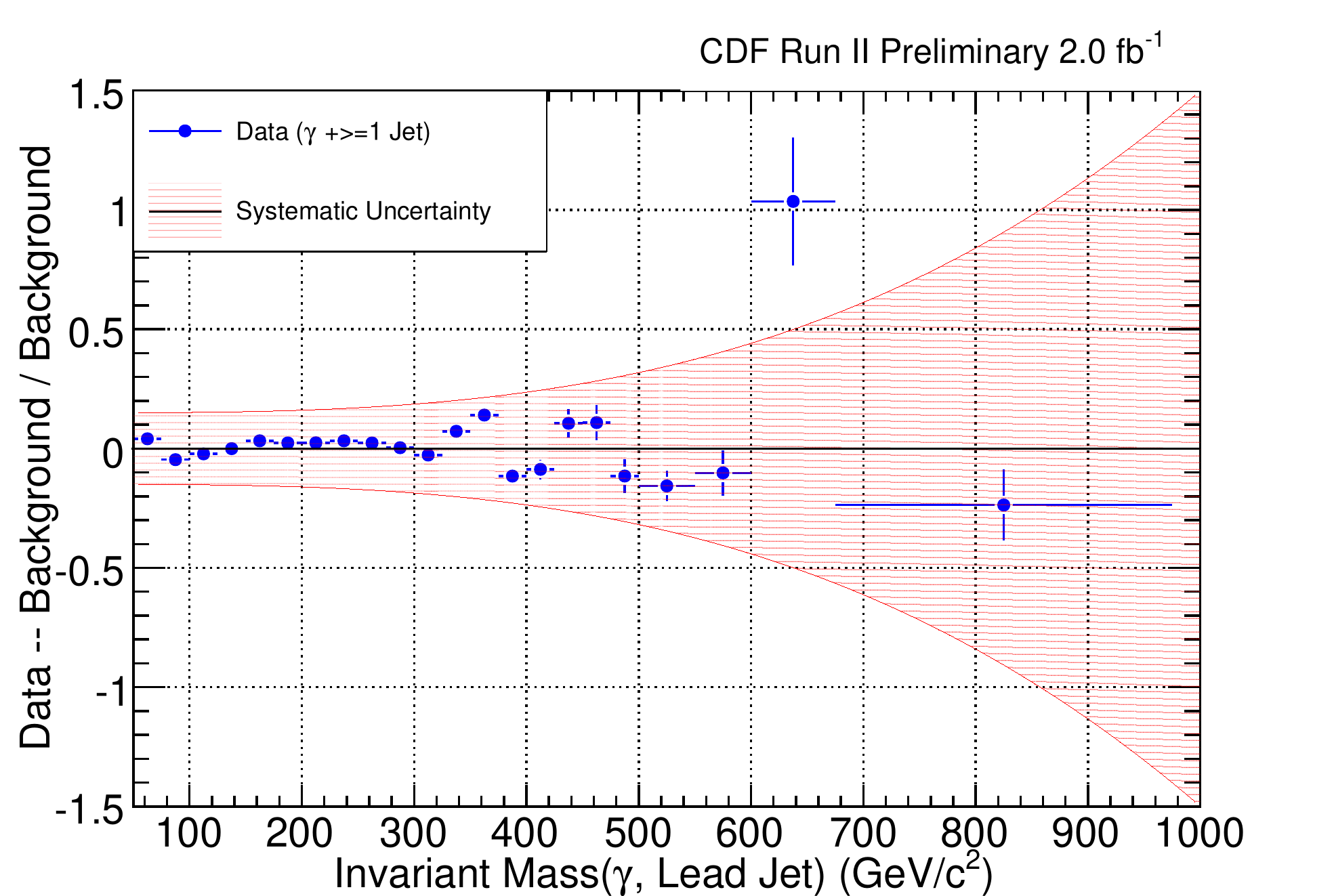}}
\caption{Photon $E_{T}$ (top) and invariant mass of the $\gamma$ + leading $E_T$ jet (bottom) in \phoonejet events.}
\end{figure*}


\end{document}